\documentstyle[citesort,12pt]{article}

\def\lsim{\;\raise0.3ex\hbox{$<$\kern-0.75em\raise-1.1ex\hbox{$\sim$}}\;}
\def\gsim{\;\raise0.3ex\hbox{$>$\kern-0.75em\raise-1.1ex\hbox{$\sim$}}\;}
 \newcommand{\ba}{\begin{array}}
\newcommand{\ea}{\end{array}} \newcommand{\bea}{\begin{eqnarray}}
\newcommand{\eea}{\end{eqnarray}} 

\begin{document}

\begin{titlepage}

\begin{flushright}
Orsay LPTHE-99-03
\end{flushright}

\vspace{3cm}
\centerline{\Large\bf Cascade Decays in the NMSSM
\footnote{To appear in the proceedings of the BTMSSM subgroup of the
Physics at Run II Workshop on Supersymmetry/Higgs}}

\vspace{2cm}
\begin{center}
{\bf U. Ellwanger and C. Hugonie}\\
Laboratoire de Physique Th\'eorique et Hautes Energies
\footnote{Laboratoire associ\'e au Centre National de la Recherche Scientifique
(URA D0063)}\\    
Universit\'e de Paris XI, Centre d'Orsay, F-91405 Orsay Cedex, France\\  
\end{center}
\vspace{2cm}

\begin{abstract}

We study unconventional signatures of the NMSSM (the MSSM with an additional
gauge singlet) with a singlino LSP. Compared to sparticle production processes
in the MSSM, these consist in additional cascades 
(one or two additional $l^+ l^-$, $\tau^+ \tau^-$ or $b \bar b$ pairs or 
photons), possibly with macroscopically displaced vertices with distances 
varying from millimeters to several meters.

\end{abstract}

\end{titlepage}

\noindent
{\bf Definition of the Model:}
\vspace{.4 cm}

The NMSSM (Next-to-minimal SSM, or (M+1)SSM) is defined by the addition of
a gauge singlet superfield $S$ to the MSSM. The superpotential $W$ is scale
invariant, i.e. there is no $\mu$-term. Instead, two Yukawa couplings
$\lambda$ and $\kappa$ appear in $W$. Apart from the standard quark and
lepton Yukawa couplings, $W$ is given by
\bea W = \lambda H_1 H_2 S + \frac{1}{3} \kappa S^3 + \ldots \label{sp} \eea
and the corresponding trilinear couplings $A_{\lambda}$ and $A_{\kappa}$
are added to the soft susy breaking terms. The vev of $S$ generates an
effective $\mu$-term with $\mu = \lambda \langle S \rangle $.

The constraint NMSSM (CNMSSM) [1] is defined by universal soft susy breaking
gaugino masses $M_0$, scalar masses $m_0^2$ and trilinear couplings
$A_0$ at the GUT scale, and a number of phenomenological constraints:

\noindent
- Consistency of the low energy spectrum and couplings with negative
Higgs and sparticle searches.

\noindent
- In the Higgs sector, the minimum of the effective potential with 
$\langle H_1 \rangle $ and $ \langle H_2 \rangle \neq 0$ has to be 
deeper than any minimum with $ \langle H_1 \rangle $ and/or 
$ \langle H_2 \rangle = 0$. Charge and colour breaking minima induced by
trilinear couplings have to be absent. (However, deeper charge and 
colour breaking minima in "UFB" directions are allowed, since the
decay rate of the physical vacuum into these minima is usually large
compared to the age of the universe [2].)

Cosmological constraints as the correct amount of dark matter are not
imposed at present. (A possible domain wall problem due to the discrete
$Z_3$ symmetry of the model is assumed to be solved by, e.g., embedding
the $Z_3$ symmetry into a $U(1)$ gauge symmetry at $M_{GUT}$, or by
adding non-renormalisable interactions which break the $Z_3$ symmetry
without spoiling the quantum stability [3].)

The number of free parameters of the CNMSSM, ($M_{1/2}$, $m_0$, $A_0$,
$\lambda$, $\kappa$ + standard Yukawa couplings), is the same as in the
CMSSM ($M_{1/2}$, $m_0$, $A_0$, $\mu$, $B + \ idem$). The new physical
states in the CNMSSM are one additional neutral Higgs scalar and Higgs
pseudoscalar, respectively, and one additional neutralino. In general
these states mix with the corresponding ones of the MSSM with a mixing
angle proportional to the Yukawa coupling $\lambda$. However, in the
CNMSSM $\lambda$ turns out to be quite small, $\lambda \lsim 0.1$ (and
$\lambda \ll 1$ for most allowed points in the parameter space) [1]. Thus
the new physical states are generally almost pure gauge singlets with
very small couplings to the standard sector.

\vspace{.5 cm}
\noindent
{\bf Phenomenology of the CNMSSM:}
\vspace{.5 cm}

The new states in the Higgs sector can be very light, a few GeV or
less, depending on $\lambda$ [4]. Due to their small couplings to the $Z$
boson they will escape detection at LEP and elsewhere, i.e. the lightest
``visible'' Higgs boson is possibly the next-to-lightest Higgs of the
NMSSM. The upper limits on the mass of this visible Higgs boson (and its
couplings) are, on the other hand, very close to the ones of the MSSM,
i.e. $\lsim 140$ GeV depending on the stop masses [4].

The phenomenology of sparticle production in the CNMSSM can differ
considerably from the MSSM, depending on the mass of the additional
state $\tilde S$ in the neutralino sector: If the $\tilde S$ is not the
LSP, it will hardly be produced, and all sparticle decays proceed as in
the MSSM with a LSP in the final state. If, on the other
hand, the $\tilde S$ is the LSP, the sparticle decays will proceed
differently: First, the sparticles will decay into the NLSP, 
because the couplings to the $\tilde S$ are too small. Only
then the NLSP will realize that it is not the true LSP, and decay
into the $\tilde S$ plus an additional cascade. 

The condition for a singlino LSP scenario can be expressed relatively
easily in terms of the bare parameters of the CNMSSM: Within the allowed 
parameter space of the CNMSSM, the lightest non-singlet neutralino is 
essentially a bino $\tilde B$. Since the masses of $\tilde S$ and $\tilde B$
are proportional to $A_0$ and $M_{1/2}$, respectively, one finds,
to a good approximation, that the $\tilde S$ is the true LSP if the bare
susy breaking parameters satisfy $|A_0| \lsim 0.4 M_{1/2}$. Since $A_0^2
\gsim 9 m_0^2$ is also a necessary condition within the CNMSSM, the
singlino LSP scenario corresponds essentially to the case where the
gaugino masses are the dominant soft susy breaking terms.

Note, however, that the $\tilde B$ is not necessarily the NLSP in this case:
Possibly the lightest stau $\tilde\tau_1$ is lighter than the $\tilde B$,
since the lightest stau can be considerably lighter than the sleptons of the 
first two generations. Nevertheless, most sparticle decays will proceed via
the $\tilde B \to \tilde S + \ldots$ transition, which will give rise to
additional cascades with respect to decays in the MSSM. The properties of 
this cascade have been analysed in [5], and in the following we will
briefly discuss the branching ratios and the $\tilde B$ life times in
the different parameter regimes:

a) $\tilde B\to \tilde S \nu\bar\nu$: This invisible process is mediated
dominantly by sneutrino exchange. Since the sneutrino mass, as the mass of
$\tilde B$, is essentially fixed by $M_{1/2}$ [5], the
associated branching ratio varies in a predictable way with $M_{\tilde B}$: It
can become up to 90\% for $M_{\tilde B} \sim 30$~GeV, but decreases with
$M_{\tilde B}$ and is maximally 10\% for $M_{\tilde B} \gsim 65$~GeV. 

b) $\tilde B \to \tilde S l^+l^-$: This process is mediated dominantly by the
exchange of a charged slepton in the s-channel. If the lightest stau
$\tilde\tau_1$ is considerably lighter than the sleptons of the first two
generations, the percentage of taus among the charged leptons can well exceed
$\frac{1}{3}$. If $\tilde\tau_1$ is lighter than $\tilde B$, it is produced
on-shell, and the process becomes $\tilde B \to \tilde\tau_1 \tau \to \tilde S
\tau^+ \tau^-$. Hence we can have up to 100\% taus among the charged leptons and
the branching ratio of this channel can become up to 100\%. 

c) $\tilde B\to \tilde S S$: This two-body decay is kinematically allowed if
both $\tilde S$ and $S$ are sufficiently light. (A light $S$ is not excluded by
Higgs searches at LEP1, if its coupling to the $Z$ is too
small [4].) However, the coupling $\tilde B \tilde S S$ is
proportional to $\lambda^2$, whereas the couplings appearing in the decays a)
and b) are only of $O(\lambda)$. Thus this decay can only be important for
$\lambda$ not too small. In [5], we found that its branching ratio can
become up to 100\% in a window $10^{-3} \lsim \lambda \lsim 10^{-2}$. Of
course, $S$ will decay immediately into $b\bar b$ or $\tau^+ \tau^-$, depending
on its mass. (If the branching ratio $Br(\tilde B\to \tilde S S)$ is
substantial, $S$ is never lighter than $\sim 5$~GeV.) If the singlet is heavy 
enough, its $b\bar b$ 
decay gives rise to 2 jets with $B$ mesons, which are easily detected with
$b$-tagging. In any case, the invariant mass of the $b\bar b$  or the 
$\tau^+ \tau^-$ system would be peaked at $M_S$, making
this signature easy to search for.

d) $\tilde B\to \tilde S \gamma$: This branching ratio can be important if the
mass difference $\Delta M = M_{\tilde B} - M_{\tilde S}$ is small ($\lsim
5$~GeV). 

Further possible final states like $\tilde B \to \tilde S q\bar q$ via $Z$
exchange have always branching ratios below 10\%. (The two-body decay $\tilde B
\to \tilde S Z$ is never important, even if $\Delta M$ is larger than $M_Z$: In
this region of the parameter space $\tilde\tau_1$ is always the NLSP, and thus
the channel $\tilde B \to \tilde\tau_1 \tau$ is always prefered.)

The $\tilde B$ life time depends strongly on the Yukawa coupling $\lambda$,
since the mixing of the singlino $\tilde S$ with gauginos and higgsinos is
proportional to $\lambda$. Hence, for small $\lambda$ (or a small mass 
difference $\Delta M$) the $\tilde B$ can 
be so long lived that it decays only after a macroscopic lenght of flight 
$l_{\tilde B}$. An approximate formula for $l_{\tilde B}$ (in meters) is given 
by 
\bea l_{\tilde B}[m] \simeq 2\cdot 10^{-10} \frac{1}{\lambda^2 \cdot
M_{\tilde B}[GeV]}\ ,\eea
and $l_{\tilde B}$ becomes $>\ 1$ mm for $\lambda \lsim 6\cdot 10^{-5}$.

\vspace{.4cm}
To summarize, the following unconventional signatures are possible
within the CNMSSM, compared to the MSSM:

\noindent
a) additional cascades attached to the original vertex (but still
missing energy and momentum): one or two additional $l^+ l^-$, $\tau^+ \tau^-$
or $b \bar b$ pairs or photons, with the corresponding branching ratios
depending on the parameters of the model.

\noindent
b) one or two additional $l^+ l^-$ or $\tau^+ \tau^-$ pairs or photons with
macroscopically displaced vertices, with distances varying from
millimeters to several meters. These displaced vertices do not point
towards the interaction point, since an additional invisible particle is
produced.

More details on the allowed branching ratios and life times can be found
in [5], applications to sparticle production processes et LEP 2 are
published in [6], and differential (spin averaged) cross
sections of the $\tilde B \to \tilde S$ decay are available upon request. 

\vspace{.4 cm}
\noindent
{\bf References}
\vspace{.4 cm}

\noindent
[1] U. Ellwanger, M. Rausch de Traubenberg, C. Savoy, Nucl. Phys. 
{\bf B 492} (1997) 21

\noindent
[2] U. Ellwanger, C. Hugonie, "Constraints from Charge and Colour Breaking
Minima in the (M+1)SSM", in preparation

\noindent
[3] C. Panagiotakopoulos, K. Tamvakis, hep-ph/9809475\newline
\indent S.A. Abel, {\em Nucl. Phys.} {\bf B 480} (1996) 55

\noindent 
[4] U. Ellwanger, M. Rausch de Traubenberg, C. Savoy, Z. Phys. 
{\bf C 67} (1995) 665

\noindent
[5] U. Ellwanger, C. Hugonie, Eur. Phys. J. {\bf C5}
(1998) 723

\noindent
[6] U. Ellwanger, C. Hugonie, hep-ph/9812427

\end{document}